\documentstyle[aps,multicol,epsf]{revtex}
\begin{document}
\draft
\title{Half-quantum vortex and ${\hat d}$-soliton in Sr$_2$RuO$_4$ }

\author{Hae-Young Kee$^{1}$, Yong Baek Kim$^{2}$, and Kazumi Maki$^{3}$}
\address{
$^1$ Department of Physics, University of California,
Los Angeles, CA 90095\\
$^2$ Department of Physics, The Ohio State University,
Columbus, OH 43210\\
$^3$ Department of Physics,
University of Southern California, Los Angeles, CA 90089}

\date{\today}
\maketitle

\begin{abstract}
Assuming that the superconductivity in Sr$_2$RuO$_4$ is described by
a planar p-wave order parameter, we consider possible topological 
defects in Sr$_2$RuO$_4$.
In particular, it is shown that both of the ${\hat d}$-soliton and 
half-quantum vortex can be created in the presence of  
the magnetic field parallel to the $a$-$b$ plane.
We discuss how one can detect the ${\hat d}$-soliton and 
half-quantum vortex experimentally.
\end{abstract}


It has been suggested that the unconventional superconducting 
state of Sr$_2$RuO$_4$ is described by the planar spin-triplet 
$p$-wave order parameter with broken time reversal symmetry
in analogy to the $^3$He A-phase.
As known from the example of superfluid $^3$He, one 
of the hallmarks of the triplet superconductivity is the
presence of a manifold of topological defects.\cite{voll,volo}
Thus, we expect that the creation and detection of 
topological defects in Sr$_2$RuO$_4$ (or spin-triplet 
superconductor) will provide further insights about the
nature of the unconventional superconducting state of
Sr$_2$RuO$_4$. In this sense, the study of the topological
defects in the planar $p$-wave superconducting state with
broken time reversal symmetry is of great interest.

More specifically, it has been proposed\cite{rice,sig1} 
that the superconducting order parameter in this system  
is described by the planar $p$-wave form written as,
\begin{equation}
\Delta_{\alpha \beta} ({\vec k}) = {\vec d}({\vec k}) \cdot 
({\vec \sigma} i \sigma_2)_{\alpha \beta}  
\label{delta}
\end{equation}
with 
\begin{equation}
{\vec d} ({\vec k}) = \Delta {\hat d} 
({\hat k}_1 \pm i {\hat k}_2) \ , 
\label{order}
\end{equation}
where $\sigma_{\mu} (\mu=1,2,3)$ are Pauli matrices and 
$\alpha, \beta$ represent spin $\uparrow$ or $\downarrow$.
Here ${\hat k}_j (j=1,2)$ represent the projection of the 
unit wave vector ${\hat k}$ along two perpendicular directions
${\hat e}_1$ and ${\hat e}_2$ in two dimensional space.  
This order parameter describes the Cooper pair state with
the zero spin projection on ${\hat d}$ and 
with the unique projection of the pair orbital angular 
momentum given by ${\hat l} ={\hat e}_1 \times {\hat e}_2$.
In Sr$_2$RuO$_4$, due to the spin-orbit coupling, 
${\hat d}$ is forced to be parallel to $\pm {\hat c}$
and ${\vec k}$ is the quasi-particle momentum 
in the $a$-$b$ plane.

Indeed the spontaneous magnetization seen by muon 
spin relaxation experiment \cite{luke} and 
flat Knight shift seen by NMR\cite{ishi}
seem to be consistent with this picture.
On the other hand, the origin of the spontaneous 
magnetization seen by muon 
spin relaxation experiment is somewhat
mysterious since we do not expect such a 
magnetization in a homogeneous system.

It is important to notice that the 
superconducting ground state described by the order parameter 
of Eq.\ref{order} is doubly degenerate.
We can designate these two ground states by
the angular momentum $l_z = \pm 1$, where 
${\hat l}$ is parallel to the $c$-axis.
Sigrist and Agterberg\cite{sig2} proposed recently 
that there will be in general a domain wall 
between $l_z=1$ and $l_z=-1$ states, which we shall 
call ${\hat l}$-soliton\cite{maki1}
in analogy to the case of superfluid $^3$He-A.
It is likely that such a soliton is magnetically 
active, so it may be an origin of the spontaneous 
magnetization seen in muon spin relaxation 
experiment\cite{luke}.
In particular, in a magnetic field $H  \ || \ {\hat c}$, 
only one of 
these degenerate states is favored \cite{agter,wang}. 
Therefore, it is possible to control ${\hat l}$-solitons 
by a magnetic field parallel to the $c$-axis. \cite{sig2}
They also proposed that these ${\hat l}$-solitons
would provide very efficient barriers for the vortex motion
and this effect is possibly related to the pinning of 
vortices observed in Sr$_2$RuO$_4$ below $T=30mK$ \cite{mota}.
However, in this experiment the magnetic field is 
applied in a direction perpendicular to the $c$-axis.
As we will see later, so-called ${\hat d}$-solitons appear 
to be more appropriate than ${\hat l}$-solitons in this 
configuration.
Also, for the ${\hat l}$-solitons, it is rather
difficult to estimate the soliton energy and to make
a further quantitative prediction.

The purpose of this paper is to propose an alternative 
model for the appearance of the spontaneous magnetization
and the mechanism of the pinning of vortices; 
${\hat d}$-soliton and half-quantum vortex.
The ${\hat d}$-soliton is a domain wall between 
${\hat d} \parallel {\hat c}$
and ${\hat d} \parallel -{\hat c}$ as in 
superfluid $^{3}$He-A.
We believe that ${\hat d}$ is parallel and antiparallel
to ${\hat c}$, because they are forced to be parallel
to the angular momentum ${\hat l}$ (or $-{\hat l}$)
due to the spin-orbit coupling characterized
by an energy scale $\Omega_d$ \cite{kee}.
Therefore, if we use $\Omega_d$ as a parameter, we can 
calculate the energy and shape of the ${\hat d}$-soliton
provided that $\Omega_d \ll \Delta(T)$, where $\Delta (T)$ 
is the superconducting gap. 
Unfortunately we do not know the precise value 
of $\Omega_d$, but it may be about ${1 \over 10}\Delta(T)$. 
In this picture, moving ${\hat d}$-soliton generates
the local magnetization which can result in the spontaneous
magnetization seen in muon spin relaxation experiment.
One can also generate a large number of ${\hat d}$-solitons
by applying a burst of high frequency microwave with 
frequency $\sim \Omega_d$ sent parallel to the 
$a$-$b$ plane.\cite{maki1}

As in superfluid $^3$He-A, each ${\hat d}$-soliton is 
terminated by a pair of half-quantum vortices.\cite{salo}
We find that these pairs of half-quantum vortices are more
stable than the usual single quantum vortex in the 
superconducting state in the presence of the magnetic 
field parallel to the $a$-$b$ plane. 
This means that the usual single quantum vortex
would spilt into a pair of half-quantum vortices 
connected by the ${\hat d}$-soliton. \cite{volo,maki2}
In this case, these objects would provide an extremely 
efficient pinning mechanism of vortices in Sr$_2$RuO$_4$.
Also the half-quantum vortices should exhibit
a clear electron spin resonance (ESR) signature.
Further we believe that these objects are visible by
the scanning tunneling microscopy (STM) imaging and 
by micromagnetometry developed by Kirtley 
{\it et al}\cite{kirt} 
used in high $T_c$ cuprate compounds.

\vskip 0.3cm
{\bf Free energy of the conventional single vortex when the 
magnetic field is parallel to the $a$-$b$ plane}

Let us assume that the magnetic field is
parallel to the $a$-axis.
Then the free energy of the conventional vortex with the
flux quantum $\phi_0 = hc/2e$ is obtained within
the London approximations as
\begin{equation}
f_v=(\frac{\phi_0}{4\pi \lambda})^2 {\rm ln} (\frac{\lambda}{\xi}) \ ,
\end{equation}
where $\xi$ is the coherence length and $\lambda$ is the magnetic
penetration depth.
The magnetic penetration depth $\lambda$ is related to the
superfluid density $\rho_s(T)$ by
$\lambda^{-2}=\frac{4\pi e^2}{m c^2} \rho_s(T)$.
When the magnetic field is parallel to the $a$-axis,
in the anisotropic system like Sr$_2$RuO$_4$, 
$\lambda$ and $\xi$ should be reinterpreted as 
$\lambda=\sqrt{\lambda_b \lambda_c}$
and $\xi=\sqrt{\xi_b \xi_c}$. Here $\lambda_{b,c}$
and $\xi_{b,c}$ are the magnetic penetration depth
and coherence length in the $b$ and $c$ directions
respectively. 

\vskip 0.3cm
{\bf ${\hat d}$-soliton and a pair of half-quantum vortices}

There exists huge anisotropy in the in-plane and out-of-plane
transport properties in Sr$_2$RuO$_4$. Thus Sr$_2$RuO$_4$ may 
be regarded as an effectively two dimensional system.
The large anisotropy or the effective two-dimensionality
of the system forces the angular momentum of the Cooper
pair to be parallel or antiparallel to the $c$-axis.
In the $p$-wave superconducting state described by the 
order parameter described by Eq.\ref{delta} and 
Eq.\ref{order}, the ${\hat d}$ vector 
is oriented along $\pm {\hat l}$ in the presence of the
spin-orbit coupling.
Here we consider the case that the angular momentum ${\hat l}$ 
is uniform in the entire system. We can assume, without loss
of generality, that ${\hat l} \parallel {\hat c}$.
We are interested in the deformation of the ${\hat d}$ 
configuration from the uniform case; for example,
${\hat d} \parallel {\hat c}$.  
Any deviation from the uniform state would cost the
energy associated with the spin-orbit coupling 
characterized by an energy scale $\Omega_d$ \cite{kee}.
However, we will show that the so-called ${\hat d}$-soliton
(a particular form of the ${\hat d}$ configuration)
with a pair of half-quantum vortices can have lower energy 
than the conventional single vortex.   
Thus it is easier to excite a pair of half-quantum vortices
with a ${\hat d}$-soliton compared to single conventional 
vortex.
In particular, a magnetic field 
parallel to the $a$-$b$ plane generates very likely 
pairs of half-quantum vortices rather than usual 
vortices when the formers are stable.

We consider the ${\hat d}$-soliton that is a topological
planar defect in the ${\hat d}$ configuration.
The orientation of ${\hat d}$ changes by $\pi$ across 
the planar defect while ${\hat d}$ vectors at far
distances are still along the $c$-axis. 
Typical configurations of ${\hat d}$-soliton in the
$y$-$z$ plane can be  
found in Fig.1 and Fig.2 which we will explain later.    
We take $y$ and $z$ as the coordinates along $b$-axis and
$c$-axis respectively.
Now let us attach a pair of half-quantum vortices to the 
end points of the ${\hat d}$-soliton of length $R$ in 
$y$-$z$ plane.
In the case of an isolated half-quantum vortex, we have
$e^{i\pi}=-1$ factor in the order parameter due to 
phase winding around the half-quantum vortex. 
Therefore, an isolated half-quantum vortex cannot occur.
On the other hand, if the half-quantum vortex is 
attached to the end points of the ${\hat d}$-soliton,
the disgyration in ${\hat d}$ at the same point
compensates the phase $\pi$ so that there is no net
change in the overall phase of the order parameter. 

In order to show that a pair of half-quantum vortices
with the ${\hat d}$-soliton is a lower energy excitation
compared to single conventional vortex, we have to
compare the free energies of two cases.  
The free energy required to create the ${\hat d}$-soliton 
is obtained from 
\begin{equation}
f_d=\frac{1}{2} \chi_N C^2 \int d^3 r [\sum_{i j} |\partial_i {\hat d}_j|^2
+\xi_d^{-2} (1-d_z^2) ] \ ,
\label{dsoliton}
\end{equation}
where $\chi_N$ is the spin susceptibility, 
$\xi_d(T) =C(T)/\Omega_d(T)$ where $C(T)$ is the spin wave
velocity, and $\Omega_d(T)$ is the longitudinal spin resonance frequency.
\cite{kee}

On the other hand, ${\hat d}$ vector of the ${\hat d}$-soliton can 
be parametrized by the following expression.
\begin{equation}
{\hat d}=\cos \psi {\hat z}+\sin \psi {\hat y} \ ,
\end{equation}
where
\begin{equation}
\psi(y,z)=\frac{1}{2} \left(\arctan{\frac{z+R/2}{y}} - \arctan{\frac{z-R/2}{y}}
\right) \ ,
\label{zsoliton}
\end{equation}
where we put two half-quantum vortices at $(y,z)=(0,R/2)$ and
$(0,-R/2)$.

In the past, similar form of $\psi$ was also discussed in a different
context in regard to $^{3}$He. \cite{salo}
As one can see, there is a discontinuity in $\psi$ 
across the line defined by $-R/2 < z < R/2$ and $y=0$.
The spatial configuration of the corresponding ${\hat d}$ 
around a pair of half-quantum vortices is shown in Fig.1.
As one can see from the figure, the planar defect is parallel to 
the ${\hat z}$-direction or $c$-axis.
One can also consider the planar defect lying along the ${\hat y}$-axis
given by 
\begin{equation}
\psi=\frac{1}{2} \left( \arctan{\frac{y+R/2}{z}} - \arctan{\frac{y-R/2}{z}}
\right) \ , 
\label{ysoliton}
\end{equation}
where two half-quantum vortices are located at $(y,z)=(R/2,0)$
and $(-R/2,0)$.
The configuration of the ${\hat d}$ vector using the above 
$\psi$ is shown in Fig.2.
One can easily see that the free energies, $f_d$, associated with 
two possible ${\hat d}$ configurations are the same. 

The total free energy of the ${\hat d}$-soliton and a pair
of half-quantum vortices is given by
\begin{eqnarray}
f_{pair}&=&\frac{1}{2} \chi_N C^2 \int dy dz \ [ K (\nabla \Phi)^2
i+ \sum_{i j} |\partial_i {\hat d}_j|^2
+\xi_d^{-2} \sin^2 \psi ]
\nonumber\\
&=& \frac{1}{2} \chi_N C^2 ( \pi K {\rm ln}{\frac{\lambda}{R}}
+I_1+I_2 ) \ ,
\label{free}
\end{eqnarray}
where $\Phi$ represents the phase of the order parameter
which couples to the external electromagnetic field.
The parameter $K$ is defined by
\begin{equation}
K=\frac{\rho_s}{\rho_{sp}}
=\frac{1+1/3 F_1}{1+1/3 F_1^a} \frac{1+1/3 F_1^a(1-\rho_s^0)}
{1+1/3 F_1(1-\rho_s^0)} \ ,
\label{K}
\end{equation}
where $\rho_s$ and $\rho_{sp}$ are the superfluid density
and the spin superfluid density respectively.
$F_1$ and $F_1^a$ are the Landau Parameters
and $\rho_s^0$ ($\equiv 1-Y(T)$ and $Y(T)$ is the
Yosida function) is the superfluid density without the 
Fermi liquid correction. \cite{voll}
Notice that $K(T_c)=1$ at $T=T_c$ and 
$K(0)=\frac{1+1/3F_1}{1+1/3 F^a_1}$ at $T=0$.
The temperature dependence of the parameter $K$ 
is shown in Fig.3 assuming that 
$F_1=9$ and $F_1^a=0$. This choice of the parameters
will be explained later.

The first term in the first and second lines of 
Eq.\ref{free} is the contribution from two 
half-quantum vortices and represents the fact that
these half-quantum vortices repel each other. 
$I_{1,2}$ are the contributions from the second 
and third terms in the first line of Eq.\ref{free}.
These contributions come from the disclination of 
the ${\hat d}$-vector.

Using the form of $\psi(y,z)$ discussed above, 
Eq.\ref{zsoliton}, $I_1$ and $I_2$ can be obtained as follows.
\begin{eqnarray}
I_1 &= & \frac{1}{4} \int dy dz \frac{R^2}{[y^2+(z+R/2)^2][y^2+(z-R/2)^2]}
\nonumber\\
&=& \pi {\rm ln}{\frac{R}{\xi}},
\nonumber\\
I_2 &=& \frac{1}{2 \xi_d^2} \int dy dz \left( 1-\frac{y^2+z^2-R^2/4}
{\sqrt{(y^2+z^2-R^2/4)^2+y^2 R^2}} \right)
\nonumber\\
&=& \pi \left( \frac{R}{2 \xi_d} \right)^2 
{\rm ln}{\frac{4 \xi_d}{R}},
\end{eqnarray}
where $\xi_d$ is the length scale associated with the spin-orbit
coupling defined by $\xi_d(T) = C(T)/ \Omega_d(T)$.

By minimizing $f_{pair}$ with respect to $R$, we obtain
the optimal $R_0$ for the lowest free energy configuration
of a pair of half-quantum vortices and the ${\hat d}$-soliton.
The optimal $R_0$ is given by 
\begin{equation}
R^2_0 = \frac{(K-1) 2\xi_d^2}{{\rm ln}{\frac{4 \xi_d}{\sqrt{e}R_0}}}
 >  0.
\label{conditionR} 
\end{equation}
Here we have assumed that the $\xi_d > \sqrt{e} R_0 / 4$.
Notice that the half-quantum vortices with a ${\hat d}$-soliton
is possible only when $K > 1$ in order to have $R_0 > 0$.
Although we have no information about $F_1^a$, it is most likely
that $F_1^a \sim 0$.
The ratio between the effective mass and the bare mass, 
$\frac{m^*}{m}$, is
about $4$, which means that $F_1 \sim 9$.
Therefore, $K > 1$ in the superconducting
state, as one can see from Eq.\ref{K}. 
Thus this condition is always satisfied below $T_c$.
However, the existence of the solution for $R_0$ depends on
the value of $K$.
We find that the solution exists only if
$  1 < K \le 1.5 $.
For example, for $K=1.5$, $\xi_d/R = 0.85$.
Since the parameter $K$ depends on temperature as shown in Fig.3,
we find that a pair of half-quantum vortices with 
${\hat d}$-soliton
exist only for $0.78 \le T/T_c < 1$.

%{\bf Since the distance between the half vortices of a pair which is
%given by Eq.(\ref{conditionR}), we find that
%$R$ is the same order of $\xi_d$ if we assume that $\xi_d \sim 10^{-1} \mu m$.
%? }

Now the free energy of a pair of half-quantum vortices and the
${\hat d}$-soltion at the optimal $R_0$ can be obtained as
\begin{equation}
f_{pair} = \frac{1}{2} \pi \chi_N C^2 [ K {\rm ln}{\frac{\lambda}{\xi}}
+\frac{(K-1)}{2} {\rm ln}{\frac{\Lambda \xi^2}{ 2 
(K-1) \xi_d^2}} + \frac{K-1}{2} ]
\end{equation}
where $\Lambda = {\rm ln}{\frac{4 \xi_d}{\sqrt{e}R_0}}$.
In order to examine the stability of the half-quantum vortices, 
we have to compare $f_{pair}$ and the free energy of single 
vortex, $f_v$. The difference is given by
\begin{equation}
f_{v} - f_{pair}
= \frac{1}{2}\pi \chi_N C^2 [ {\rm ln}{\frac{\lambda}{\xi}}
+\frac{(K-1)}{2} {\rm ln}{\frac{2 (K-1) \xi_d^2}{\Lambda \lambda^2}}
-\frac{(K-1)}{2} ] \ .
\end{equation}
If $f_{v} - f_{pair} > 0$ for some values of $K > 1$, a pair of
half-quantum vortices are more stable than the conventional  
single vortex.
This condition can be rewritten as
\begin{equation}
{\lambda \over \xi} 
\left ( {\xi_d \over \lambda} \right )^{K-1} > 
{e^{(K-1)/2} \Lambda^{(K-1)/2} \over 
2^{(K-1)/2}(K-1)^{(K-1)/2}} \ .
\label{stable}
\end{equation}

Recalling that the solution of Eq. (\ref{conditionR})
exists if $1 < K \le 1.5$, one can investigate the
stability condition given by Eq.\ref{stable}.  
%Notice that as $K \rightarrow 1$, a pair of half quantum vortices
%can not exist because $R_0 \rightarrow 0$ from Eq. (\ref{conditionR}).
One can see from Eq.\ref{conditionR} and Eq.\ref{stable}
that, as long as $K > 1$, a pair of half-quantum vortices can be 
stabilized over single vortex under certain conditions for the 
ratio between $\xi_d$ and $\lambda$. For example, for 
$K=1.1$ and $K=1.5$, the conditions for the stability of 
a pair of half-quantum vortices over single vortex are given by
\begin{equation}
\frac{\xi_d}{\lambda} > 10^{-11} \ \ 
{\rm and} \ \ 
\frac{\xi_d}{\lambda} > 0.0094 \ , 
\end{equation}
respectively. 
One can see that these conditions are easily satisfied.
Here we use $\lambda/\xi=\sqrt{(\lambda_b \lambda_c)/(\xi_b \xi_c)}
=12.186$ which is appropriate for Sr$_2$RuO$_4$.
One can also see that the stability of a pair of half-quantum vortices
with the ${\hat d}$-soliton is determined by the value of $K$ 
which depends on temperatures as shown in Fig.3.

Now let us discuss the relation between 
${\hat l}$- and ${\hat d}$-solitons.
It is difficult to estimate the energy of ${\hat l}$-soliton
in terms of the texture free energy given by Eq.\ref{dsoliton}.
However, it is likely that ${\hat l}$-soliton costs much more
energy because, if it exists, the order parameter given by 
Eq.\ref{order} should vanish inside the ${\hat l}$-soliton.
Therefore, if there is a natural passage for conversion of
${\hat l}$-solitons to ${\hat d}$-solitons, most of
${\hat l}$-solitons will be converted into ${\hat d}$-solitons.

In summary, assumming that the superconducting state of 
Sr$_2$RuO$_4$ is characterized by the spin-triplet order 
parameter with broken time reversal symmetry, we investigated 
the existence of half-quantum vortices and associated topological
defect; ${\hat d}$-soliton. 
We showed that a pair of half-quantum vortices attached to
a ${\hat d}$-soliton can be created in the presence of the
magnetic field parallel to the $a$-$b$ plane.
It was found that a pair of half-quantum vortices with 
a ${\hat d}$-soliton is more stable than the 
conventional single vortex for certain temperatures below $T_c$. 
As in superfluid $^3$He-A, the presence of ${\hat d}$-soliton
may be detected as the deficit in the intensity of
electron spin resonance signal at $\omega=\Omega_d$.\cite{maki1}
There should be a clear electron spin resonance signature due to 
the half-quamtum vortices. 
Detection of the half-quantum vortices by scanning
tunneling microscopy (STM) would also provide
a convincing evidence for the spin-triplet pairing 
state with time reversal symmetry breaking. 

We thank Manfred Sigrist and Ying Liu for helpful discussions.
The work of H.-Y. Kee was conducted under the auspices of the Department
of Energy, supported (in part) by funds provided by the University of
California for the conduct of discretionary research by Los Alamos
National Laboratory. This work was also supported by NSF CAREER award
grant No. DMR-9983783 (Y.B.K.) and Alfred P. Sloan Foundation (Y.B.K.).

\pagebreak
\begin{figure}
%\hspace{1.0truecm}
\vspace{-2.5truecm}
\center
\centerline{\epsfysize=4.0in
\epsfbox{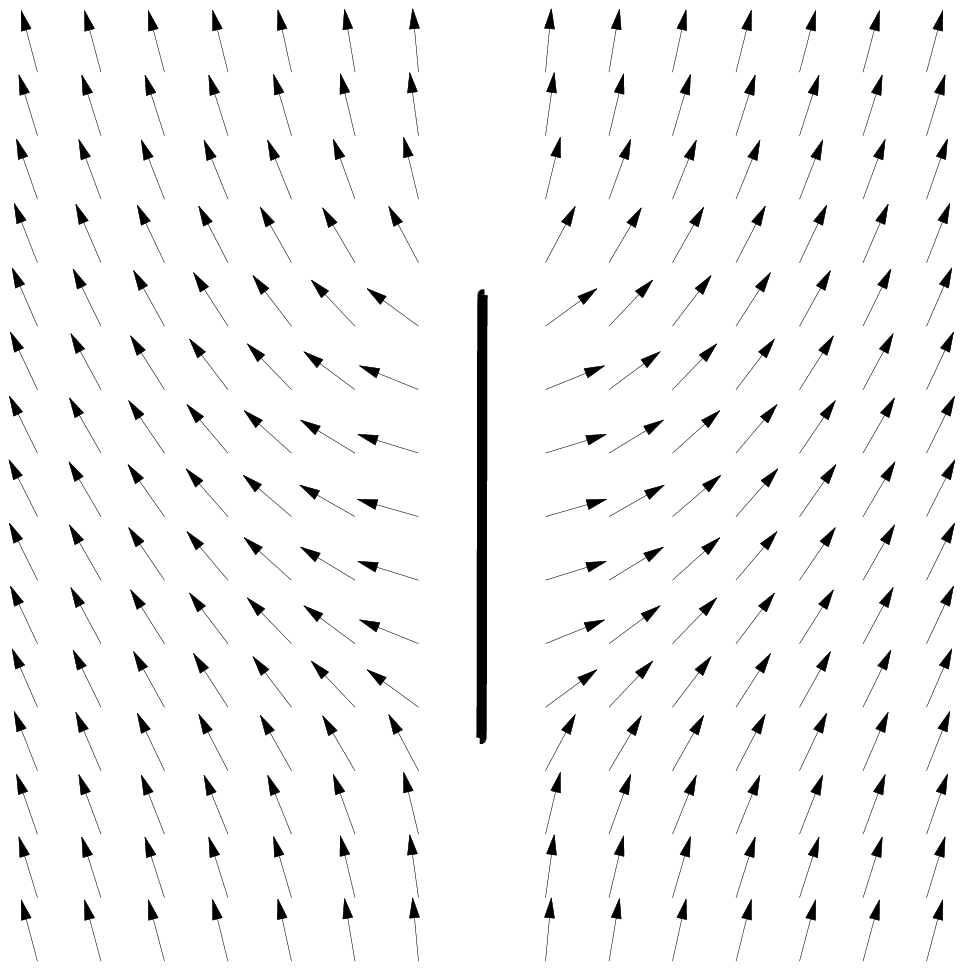}}
\vspace{1.0truecm}
\begin{minipage}[t]{10.0cm}
\caption{The spatial configuration of $d$-vector
in the $b$-$c$ plane given by Eq.\ref{zsoliton}. 
The thick line denotes the domain wall
with the length of $R$, which is parallel to 
the $c$ axis.}
\end{minipage}
\end{figure}

\pagebreak
\begin{figure}
%\hspace{1.0truecm}
\vspace{-2.5truecm}
\center
\centerline{\epsfysize=4.0in
\epsfbox{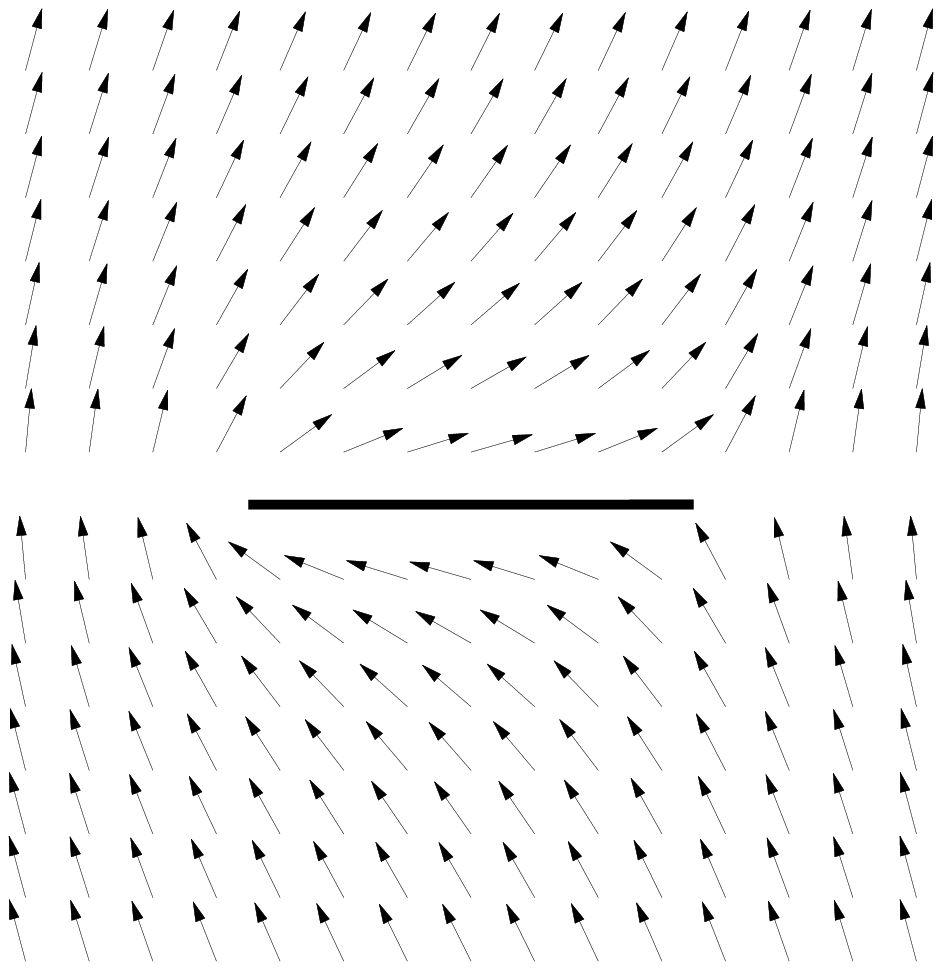}}
\vspace{1.0truecm}
\begin{minipage}[t]{10.0cm}
\caption{The spatial configuration of $d$-vector
in the $b$-$c$ plane given by Eq.\ref{ysoliton}. 
The thick line denotes the domain wall
with the length of $R$, which is parallel to the
$b$ axis.}
\end{minipage}
\end{figure}

\pagebreak
\begin{figure}
%\hspace{1.0truecm}
\vspace{-2.5truecm}
\center
\centerline{\epsfysize=3.5in
\epsfbox{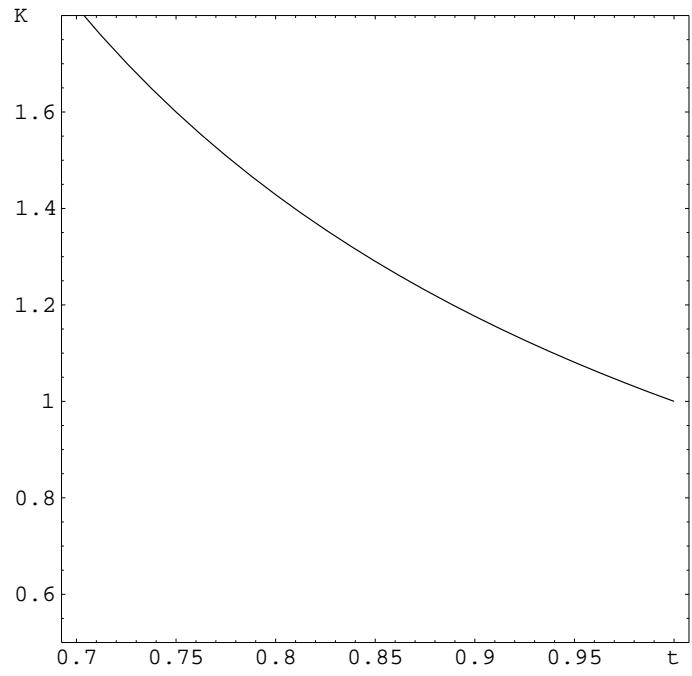}}
\vspace{1.0truecm}
\begin{minipage}[t]{10.0cm}
\caption{The parameter $K$ as a function of 
the reduced temperature $t=T/T_c$.}
\end{minipage}
\end{figure}

\end{document}